\documentclass[11pt]{article}

\usepackage{graphicx}
\usepackage{dcolumn}
\usepackage{bm}
\usepackage{amsmath}
\usepackage{latexsym}
\usepackage{amssymb}
\usepackage{authblk}

\def\fr{\frac}

\def\be{\begin{equation}}
\def\ee{\end{equation}}

\def\mR{\mathcal{R}}

\def\al{\alpha}
\def\ga{\gamma}

\def\ka{\kappa}
\def\la{\lambda}

\newcommand{\bg}[1]{{\bar{\gamma}_{#1}}}
\newcommand{\bh}[1]{{\bar{h}_{#1}}}
\def\bn{\bar{\nabla}}

\def\mn{\mu \nu}

\usepackage{color}
\input{colordvi.tex}
\begin{document}

\title{Universal instability of hairy black holes in Lovelock-Galileon theories in $D$ dimensions}

\author[1,2]{Kazufumi Takahashi}
%	\email{Email: ktakahashi@resceu.s.u-tokyo.ac.jp}
	\affil[1]{Research Center for the Early Universe (RESCEU), Graduate School of Science, The University of Tokyo, Tokyo 113-0033, Japan}
	\affil[2]{Department of Physics, Graduate School of Science, The University of Tokyo, Tokyo 113-0033, Japan}
\author[1]{Teruaki Suyama}
%	\email{Email: suyama@resceu.s.u-tokyo.ac.jp}
\author[3]{Tsutomu Kobayashi}
%	\email{Email: tsutomu@rikkyo.ac.jp}
	\affil[3]{Department of Physics, Rikkyo University, Toshima, Tokyo 175-8501, Japan}
	
\maketitle

\begin{abstract}
We analyze spherically symmetric black hole solutions with time-dependent scalar hair 
in a class of Lovelock-Galileon theories, 
which are the scalar-tensor theories with second-order field equations in arbitrary dimensions.
We first show that known black hole solutions in five dimensions
are always plagued by the ghost/gradient instability in the vicinity of the horizon.
We then generalize such black hole solutions to higher dimensions and show that
the same instability found in five dimensions appears universally in any number of dimensions.
\end{abstract}

%%%%%%%%%%%%%%%%%%%%%%%%%%%%%%%%%%%%%%%%%%%%%%%%%%%%%%%%%%%%%%%%%%%%%%%%%%%%%%%%%%%%
%%%%%%%%%%%%%%%%%%%%%%%%%%%%%%%%%%%%%%%%%%%%%%%%%%%%%%%%%%%%%%%%%%%%%%%%%%%%%%%%%%%%
%	Introduction
%%%%%%%%%%%%%%%%%%%%%%%%%%%%%%%%%%%%%%%%%%%%%%%%%%%%%%%%%%%%%%%%%%%%%%%%%%%%%%%%%%%%
%%%%%%%%%%%%%%%%%%%%%%%%%%%%%%%%%%%%%%%%%%%%%%%%%%%%%%%%%%%%%%%%%%%%%%%%%%%%%%%%%%%%
\section{Introduction}

In four-dimensional spacetime, an arbitrary linear combination of 
the Ricci scalar and a constant defines the Lagrangian density for
the unique covariant metric theory for which field
equations for the metric are of
second-order differential equations \cite{Lovelock-Rund}.
When we extend this theory by including a scalar field and still require
the second-order nature of the field equations both for the metric and the
scalar field, we obtain the theory space spanned by the four arbitrary functions
of the kinetic term of the scalar field and the scalar field itself \cite{Horndeski:1974wa,Deffayet:2009mn,Deffayet:2011gz,Kobayashi:2011nu}.
This space is now known as the Horndeski or generalized Galileon class.

One of the interesting topics to address is to find the black hole (BH) solutions in the Horndeski class
and to check if scalar hair can exist or not.
From this perspective, there is an interesting subclass in the Horndeski class 
for which the scalar field appears in the Lagrangian density only through the
contraction of the rank-2 tensor $\nabla_\mu \phi \nabla_\nu \phi$ with
the metric and the Einstein tensor.
This Lagrangian has a manifest shift symmetry $\phi \to \phi +{\rm const}$.
A no-hair theorem for general scalar-tensor theories possessing the shift symmetry
has been established in \cite{Hui:2012qt}, although some assumptions made in proving
the theorem could be violated \cite{Sotiriou:2013qea}.
Actually, for the above mentioned theories having coupling between $\nabla_\mu \phi \nabla_\nu \phi$
and the Einstein tensor, static and spherically symmetric BH solutions with nontrivial scalar profile 
have been derived in \cite{Rinaldi:2012vy,Babichev:2013cya,Anabalon:2013oea,Minamitsuji:2013ura,Bravo-Gaete:2013dca}.
Electrically charged BHs have also been found in \cite{Cisterna:2014nua}.
In particular, in \cite{Babichev:2013cya}, it was found that the theory allows the scalar field
to depend linearly on time with nontrivial radial profile in the static and spherically 
symmetric BH spacetime.
Thus, a BH can have time-dependent scalar hair.
Such a linearly time-dependent scalar field was shown to
be allowed for a wider class of theories in \cite{Kobayashi:2014eva}.  
Although the scalar field diverges at the horizon for a fixed time coordinate,
this divergent part and the term linear in time can be combined together into
the ingoing (or outgoing) Eddington-Finkelstein coordinate.
Thus, when the scalar field is expressed in terms of the Eddington-Finkelstein and 
the radial coordinate, its radial part remains finite even at the horizon.
This means that any freely infalling observer records a finite value of the scalar field
even at the horizon and nothing singular happens there.
In this sense, the scalar field is regular on the horizon.
In addition to this interesting property,
this theory has the following intriguing mathematical property:
the derivative of the scalar field $\nabla_\mu \phi \nabla_\nu \phi$ couples to
the tensors derived from the variation of the constant and the Ricci scalar,
which are the only quantities
in the Lagrangian density
giving second-order field equations for the metric.
In other words, there is a well defined one-to-one correspondence between the
Lagrangian density involving only the metric which results in second
order field equations and the rank-2 tensor to which $\nabla_\mu \phi \nabla_\nu \phi$ couples.
In four dimensions, there are only two such pairs.
The first pair is a constant and $g_{\mu \nu}$ and the second one is the
Ricci scalar $R$ and the Einstein tensor $G_{\mu \nu}$.

This correspondence in four-dimensional spacetime can be straightforwardly 
extended to higher dimensions.
It is known that the most general Lagrangian density involving only the metric
giving rise to second-order differential equations for the metric is given by
an arbitrary linear combination of the Lovelock invariants \cite{Lovelock:1971yv}.
The number $N$ of the independent Lovelock invariants depends on the
dimension of spacetime [see Eq.~(\ref{1.4})].
Then, from the correspondence mentioned above
it is natural to consider the coupling between $\nabla_\mu\phi\nabla_\nu\phi$
and the rank-2 tensors obtained from variation of the Lovelock invariants with respect to the metric.
The Lagrangian density of the theory constructed in this way consists of $N$ 
Lovelock invariants and their $N$ counterparts.
Therefore, the theory is specified by the $2N$ free parameters.
In the rest of the paper, we call such a theory the Lovelock-Galileon theory.
It turns out that the field equations both for the metric and the scalar field are of second order \cite{Charmousis:2015txa}.

Recently, BH solutions with scalar hair for the Lovelock-Galileon theory 
have been found in five-dimensional spacetime where the Gauss-Bonnet 
combination constitutes the third Lovelock invariant \cite{Charmousis:2015txa}.
Similarly to the case in four dimensions,
the BH allows the scalar field to depend linearly on time.

Once a solution is obtained, it is quite natural to ask whether the solution 
in the Lovelock-Galileon theory is stable or not. 
As for the case in four dimensions,
this issue has been addressed recently in \cite{Ogawa:2015pea}.
Interestingly, it was found that irrespective of the choice of the parameters in 
the Lagrangian, BH spacetime with linearly time-dependent scalar hair
suffers from
ghost or gradient instabilities in the vicinity of the event horizon.
In this paper, we will show that
the solution in five-dimensional spacetime given in 
\cite{Charmousis:2015txa} has the same type of instabilities when the scalar field depends 
linearly on time. To see whether this instability is universally inherent independently of 
the dimensionality of spacetime or a special feature limited only to four and five dimensions,
we further consider BH solutions in generic Lovelock-Galileon theories in higher dimensions. 
We show that, in any number of dimensions, as is the case in four and five dimensions, 
the Lovelock-Galileon theory admits solutions for which the metric is static and spherically symmetric but the scalar field has a term linear in time in addition to the piece dependent only on the radial coordinate.
However, it turns out that our generalized solutions also have the aforementioned instability. This means that the instability is not particular to four and 
five dimensions, but arises regardless of the dimensionality of spacetime
in the Lovelock-Galileon class of scalar-tensor theories.

This paper is organized as follows. In Sec.~\ref{LG}, we introduce Lovelock-Galileon theory and see
that the field equations are of second order.
In Sec.~\ref{CT}, we review the BH solution given in \cite{Charmousis:2015txa} and
show its instability. Then, we generalize these results to
general Lovelock-Galileon theories in arbitrary dimensions in Sec.~\ref{LGBH}.
Finally, we draw our conclusions in Sec.~\ref{conclusion}.

%%%%%%%%%%%%%%%%%%%%%%%%%%%%%%%%%%%%%%%%%%%%%%%%%%%%%%%%%%%%%%%%%%%%%%%%%%%%%%%%%%%%
%%%%%%%%%%%%%%%%%%%%%%%%%%%%%%%%%%%%%%%%%%%%%%%%%%%%%%%%%%%%%%%%%%%%%%%%%%%%%%%%%%%%
%	Lovelock-Galileon theory
%%%%%%%%%%%%%%%%%%%%%%%%%%%%%%%%%%%%%%%%%%%%%%%%%%%%%%%%%%%%%%%%%%%%%%%%%%%%%%%%%%%%
%%%%%%%%%%%%%%%%%%%%%%%%%%%%%%%%%%%%%%%%%%%%%%%%%%%%%%%%%%%%%%%%%%%%%%%%%%%%%%%%%%%%
\section{Lovelock-Galileon theory}\label{LG}

Lovelock theory \cite{Lovelock:1971yv} is one of the natural extensions of
general relativity in arbitrary dimensions, whose action is written as
	\be
	S_{\rm L}\equiv \int d^Dx\sqrt{-g}\sum _{n=0}^Ma_n\mR ^{(n)}, \label{1.1}
	\ee
where $D\ge 4$ is the spacetime dimension, $a_n$'s are constants, and
	\be
	\mR ^{(n)}\equiv \fr{1}{2^n}\delta ^{\al _1\al _2\cdots \al _{2n-1}\al _{2n}}_{\, \beta _1\beta _2\, \cdots \beta _{2n-1}\beta _{2n}}R^{\beta _1\beta _2}_{\al _1\al _2}\cdots R^{\beta _{2n-1}\beta _{2n}}_{\al _{2n-1}\al _{2n}} \label{1.2}
	\ee
is the Lovelock invariant of $n$th order. Here $\delta^{\alpha_1\cdots}_{\,\beta_1\cdots}$
denotes the generalized Kronecker delta:
	\be
	\delta ^{\al _1\al _2\cdots \al _{p}}_{\, \beta _1\beta _2\, \cdots \beta _{p}}\equiv p!\, \delta ^{\, \al _1}_{[\beta _1}\delta ^{\al _2}_{\beta _2}\cdots \delta ^{\al _p}_{\beta _p]}. \label{1.3}
	\ee
One can easily find that $\mR ^{(1)}$ is nothing but the Ricci scalar. For $n=0$, it is natural to define $\mR ^{(0)}\equiv 1$, so that it serves as a cosmological constant. For $2n>D$, the Lovelock invariants are identical to zero since there exists at least one spacetime coordinate that appears more than once in the indices of the generalized delta. Note also that, for $2n=D$, $\mR^{(n)}$ is a topological invariant and so it does not contribute to the equation of motion. We therefore take
	\be
	M=\left\lfloor \fr{D-1}{2}\right\rfloor ,\label{1.4}
	\ee
where $\lfloor \, \cdot \, \rfloor$ is the floor function, and thus there is only a finite number $N=M+1$ of the independent Lovelock invariants. So, it follows that
Lovelock theory reduces to general relativity with a cosmological constant in four dimensions. Varying the action \eqref{1.1} with respect to the metric gives us the following equation of motion:
	\be
	\sum _{n=0}^Ma_nH^{(n)}_{\mn}=0, \label{1.5a}
	\ee
where $H^{(n)}_{\mn}$ is the Lovelock tensor of $n$th order:
	\be
	H^{(n)\mu}{}_\nu \equiv -\fr{1}{2^{n+1}}\delta ^{\mu \al _1\al _2\cdots \al _{2n-1}\al _{2n}}_{\, \nu \beta _1\beta _2\, \cdots \beta _{2n-1}\beta _{2n}}R^{\beta _1\beta _2}_{\al _1\al _2}\cdots R^{\beta _{2n-1}\beta _{2n}}_{\al _{2n-1}\al _{2n}}. \label{1.5b}
	\ee
For $n=0$ it is appropriate to define $H^{(0)\mu}{}_{\nu}\equiv -(1/2)\delta^\mu_{\,\nu}$.
Here terms in which covariant derivatives act on the Riemann tensors do not appear by virtue of the Bianchi identity.
Thus, Lovelock theory has second-order field equations.

Now we are ready to discuss Lovelock-Galileon theory \cite{Charmousis:2015txa}. In this theory, a scalar field $\phi$ is introduced and the following action is considered:
	\begin{align}
	S&\equiv S_{\rm L}+S_{\rm LG}, \label{1.6b} \\
	S_{\rm LG}&\equiv \int d^Dx\sqrt{-g}\sum _{n=0}^Mb_nH^{(n)\al \beta}\phi _{;\al}\phi _{;\beta}, \label{1.6}
	\end{align}
where $b_n$'s are constants.
The action is invariant under the shift of the scalar field: $\phi\to\phi+c$,
where $c$ is a constant.
The equations of motion are written as follows:
	\begin{align}
	\sum _{n=0}^M\left( a_nH^{(n)}_{\mn}+b_nE^{(n)}_{\mn}\right) &=0, \label{1.7} \\
	\sum _{n=0}^Mb_nJ^{(n)\al}{}_{;\al}&=0, \label{1.8}
	\end{align}
where
	\begin{align}
	E^{(n)}_{\mn}&\equiv -\fr{1}{2}g_{\mn}H^{(n)\al \beta}\phi _{;\al}\phi _{;\beta}+H^{(n)\al} {}_{(\mu}\phi _{;\nu )}\phi _{;\al} \nonumber \\
	&\qquad -\fr{n}{2^{n+1}}g_{\la (\mu}\delta ^{\al _1\al _2\cdots \al _{2n-1}\al _{2n} \al}_{\, \nu )\, \beta _2\cdots \beta _{2n-1}\beta _{2n}\beta}R^{\la \beta _2}_{\al _1\al _2}R^{\beta _3\beta _4}_{\al _3\al _4}\cdots R^{\beta _{2n-1}\beta _{2n}}_{\al _{2n-1}\al _{2n}}\phi _{;\al}\phi ^{;\beta} \nonumber \\
	&\qquad -\fr{n}{2^{n+1}}g_{\al _1(\mu}\delta ^{\al _1\al _2\cdots \al _{2n-1}\al _{2n} \al}_{\, \nu )\, \beta _2\cdots \beta _{2n-1}\beta _{2n}\beta}R^{\beta _3\beta _4}_{\al _3\al _4}\cdots R^{\beta _{2n-1}\beta _{2n}}_{\al _{2n-1}\al _{2n}}R^{\beta _2\beta}_{\al _2\la}\phi _{;\al}\phi ^{;\la} \nonumber \\
	&\qquad -\fr{n}{2^n}g_{\al _1(\mu}\delta ^{\al _1\al _2\cdots \al _{2n-1}\al _{2n} \al}_{\, \nu )\, \beta _2\cdots \beta _{2n-1}\beta _{2n}\beta}R^{\beta _3\beta _4}_{\al _3\al _4}\cdots R^{\beta _{2n-1}\beta _{2n}}_{\al _{2n-1}\al _{2n}}\phi _{;\al}^{;\beta _2}\phi _{;\al _2}^{;\beta}, \label{1.9} \\
	J^{(n)\al}&\equiv -H^{(n)\al \beta}\phi _{;\beta}. \label{1.10}
	\end{align}
Obviously, the field equations in Lovelock-Galileon theory are also of second order.\footnote{This second-order nature of the field equations is not violated even if the coefficients $a_n$ and $b_n$ are functions of $\phi$.}
This fact again follows from the Bianchi identity.

%%%%%%%%%%%%%%%%%%%%%%%%%%%%%%%%%%%%%%%%%%%%%%%%%%%%%%%%%%%%%%%%%%%%%%%%%%%%%%%%%%%%
%%%%%%%%%%%%%%%%%%%%%%%%%%%%%%%%%%%%%%%%%%%%%%%%%%%%%%%%%%%%%%%%%%%%%%%%%%%%%%%%%%%%
%	Previous work by Charmousis and Tsoukalas
%%%%%%%%%%%%%%%%%%%%%%%%%%%%%%%%%%%%%%%%%%%%%%%%%%%%%%%%%%%%%%%%%%%%%%%%%%%%%%%%%%%%
%%%%%%%%%%%%%%%%%%%%%%%%%%%%%%%%%%%%%%%%%%%%%%%%%%%%%%%%%%%%%%%%%%%%%%%%%%%%%%%%%%%%
\section{Instability of the Einstein-Gauss-Bonnet-Galileon black holes}\label{CT}
A static and spherically symmetric BH solution for the action (\ref{1.6b}) in five dimensions 
was obtained in \cite{Charmousis:2015txa}. Here we rederive the solution and discuss its stability under
a tensor perturbation.

Since we have three independent Lovelock invariants for $D=5$, we consider the following Einstein-Gauss-Bonnet-Galileon action:
	\be
	S=\int d^5x\sqrt{-g}\left[ a_0+a_1R+a_2\mR ^{(2)}-\fr{b_0}{2}\phi _{;\al}\phi ^{:\al}+b_1G^{\al \beta}\phi _{;\al}\phi _{;\beta}+b_2H^{(2)\al \beta}\phi _{;\al}\phi _{;\beta}\right] ,\label{0.0}
	\ee
where
	\begin{align}
	\mR ^{(2)}&=R^2-4R_{\al \beta}R^{\al \beta}+R_{\al \beta \ga \delta}R^{\al \beta \ga \delta}, \label{0.01} \\
	H^{(2)}_{\mn}&=-\fr{1}{2}g_{\mn}\mR ^{(2)}+2RR_{\mn}-4R_{\mu \al}R_{\nu}{}^{\al}+4R_{\mu \al \nu \beta}R^{\al \beta}+2R_{\mu \al \beta \ga}R_{\nu}{}^{\al \beta \ga}. \label{0.02}
	\end{align}
In \cite{Charmousis:2015txa}, the form of the metric was chosen as
	\be
	ds^2=-h(r)dt^2+\fr{dr^2}{f(r)}+r^2\bg{ij}dx^idx^j, \label{0.1}
	\ee
where $\bg{ij}$ represents the metric of a three-dimensional maximally symmetric space with spatial curvature $\ka$. It was also assumed that $\phi$ has the form of
	\be
	\phi (t,r)=qt+\psi (r), \label{0.2}
	\ee
with $q\;(\neq0)$
being a constant. Nonvanishing $q$ is essential for BH solutions with regular scalar hair.
Although the scalar field is time dependent, the metric can be static thanks to the shift symmetry of the theory.\footnote{If the coefficients $a_n$ or $b_n$ depend on $\phi$, the time dependence of the scalar field spoils the staticity of the metric.}

Let us write down the equations of motion. The scalar equation reads
	\be
	-\fr{b_0}{6}+b_1\left( \fr{fh'}{2rh}-F\right) +2b_2\fr{fh'}{rh}F=0, \label{0.20a}
	\ee
and the metric equation yields
	\begin{align}
	-\fr{1}{6}\left( a_0+\fr{b_0q^2}{2h}\right) +\left( a_1-\fr{b_1q^2}{2h}\right) \fr{fh'}{2rh}-\left( a_1+\fr{b_1q^2}{2h}\right) F~~~~~~~~~~~~~~~~~~~~~~~~~~~~~~~~~~~ \nonumber \\
	+\left( a_2-\fr{b_2q^2}{2h}\right) \fr{2fh'}{rh}F+f^2\psi'^2\left[ \fr{b_1}{2}\left( \fr{h'}{rh}+\fr{2}{r^2}\right) +2b_2\left( \fr{h'}{rh}F-\fr{fh'}{r^3h}\right) \right] &=0, \label{0.20b} \\
	-\fr{1}{6}\left( a_0-\fr{b_0q^2}{2h}\right) +\left( a_1-\fr{b_1q^2}{2h}\right) \left( \fr{f'}{2r}-F\right) +\left( a_2-\fr{b_2q^2}{2h}\right) \fr{2f'}{r}F~~~~~~~~~~~~~~~~~~~~~~ \nonumber \\
	+\left( b_1+4b_2F\right) \fr{f^2}{2r}\left( \psi'^2\right)'+f^2\psi'^2\left[ \left( b_1+4b_2F\right) \left( \fr{3}{4}\fr{f'}{rf}+\fr{1}{4}\fr{h'}{rh}\right) +\fr{b_1}{r^3}-\fr{2b_2}{r^3}f'\right] &=0. \label{0.20c}
	\end{align}
Here we have defined
	\be
	F(r)\equiv \fr{\ka -f(r)}{r^2}. \label{2.5}
	\ee
Since Eq.~\eqref{0.20a} is a quadratic equation with respect to $f$, one can easily express $f$ by $h$ and $h'$ as follows:
	\be
	f=\fr{6b_1rh+3(b_1r^2+4\ka b_2)h'\pm \sqrt{9(2b_1rh+(b_1r^2+4\ka b_2)h')^2-48b_2r(b_0r^2+6\ka b_1)hh'}}{24b_2h'}.
	\ee
With this relation, the square of $\psi'$ can be obtained as a functional of $h$ and $h'$ from Eq.~\eqref{0.20b}. 
Then Eq.~\eqref{0.20c} becomes an ordinary differential equation with respect to $h$ and 
can be solved in principle.
For more details about the solution, see \cite{Charmousis:2015txa}.

Now we argue that this solution is actually unstable under the tensor perturbation 
of the following form:
	\be
	\delta g_{ab}=\delta g_{ai}=0,~~~\delta g_{ij}=r^2\chi (t,r)\bh{ij}(x^k), \label{0.6}
	\ee
where $a,b=(t,r)$. Here $\chi$ represents the dynamical degree of freedom 
of the perturbation and $\bh{ij}$ are symmetric tensor spherical harmonics which 
satisfy \cite{Higuchi:1986wu}
	\be
	\bn ^k\bn _k\bh{ij}=-\ga _t\bh{ij},~~~\bn ^i\bh{ij}=0,~~~\bar{h}^i_i=0, \label{0.6a}
	\ee
where $\bn _i$ denotes a covariant derivative with respect to $\bg{ij}$, and the eigenvalue $\ga _t$ takes continuous positive numbers for $\ka =0,-1$ or discrete values $\ga _t=\ell (\ell +D-3)-2$ ($\ell =2,3,\cdots$) in $D$-dimensional spacetime for $\ka =1$. 
The scalar field $\phi$ does not have the tensor perturbation and hence we set $\delta \phi =0$.

Plugging the decomposition (\ref{0.6}) into the original field equations and
expanding them to first order in perturbation, 
we obtain the linear differential equation for $\chi$.
From the derived perturbation equation, we can construct the corresponding
Lagrangian up to an irrelevant constant factor. This overall factor can be fixed
by the comparison with the second-order action in the simple case for which
direct computation of the second-order action is relatively simple.
Since a series of these procedures is straightforward, we omit the intermediate
computations and give the final form of the second-order action:
	\be
	S^{(2)}=\int d^5x\sqrt{-\bar{g}}\left( \fr{\la _0}{2}\dot{\chi}^2-\fr{\la _1}{2}\chi'^2+\fr{\la _2}{2}\dot{\chi}\chi'-\fr{\la _3}{2}\chi ^2\right) \bar{h}^{kl}\bar{h}_{kl}, \label{0.20}
	\ee
where $\bar{g}$ denotes the determinant of the background metric
and $\lambda_0,~\cdots,~\lambda_3$ are the background-dependent coefficients.
As we will see below, among these coefficients, $\la _0$ and $\la _1$ determine 
the presence of ghost/gradient instability.
Their explicit forms are given by
	\begin{align}
	\la _0&=\fr{a_1}{2h}-a_2\fr{f'}{rh}+\fr{r^2}{h}\left[ -\fr{q^2}{h}\left( \fr{b_1}{2}-b_2\fr{fh'}{rh}\right) +X\left( \fr{b_1}{2}+b_2\fr{f'}{r}\right) +2b_2\fr{f}{r}X'\right] ,\label{0.21a} \\
	\la _1&=\fr{a_1}{2}f-a_2\fr{f^2h'}{rh}+r^2f\left[ \fr{q^2}{h}\left( \fr{b_1}{2}-b_2\fr{fh'}{rh}\right) -X\left( \fr{b_1}{2}-3b_2\fr{fh'}{rh}\right) \right]  ,\label{0.21b}
	\end{align}
where $X$ is the canonical kinetic term of the scalar field:
	\be
	X\equiv -\fr{1}{2}\phi _{;\al}\phi ^{;\al}=\fr{q^2}{2h}-\fr{f\psi'^2}{2}. \label{kin}
	\ee

To construct the Hamiltonian, let us introduce the canonical momentum conjugate to $\chi$ as
	\be
	\pi =\sqrt{-\bar{g}}\left( \la _0\dot{\chi}+\fr{\la _2}{2}\chi'\right) \bar{h}^{kl}\bar{h}_{kl}. \label{0.30}
	\ee
Then, the Hamiltonian is given by
	\be
	H=\int d^5x\sqrt{-\bar{g}}\left[ \fr{1}{2\la _0}\left( \fr{\pi}{\sqrt{-\bar{g}}\bar{h}^{kl}\bar{h}_{kl}}-\fr{\la _2}{2}\chi'\right) ^2+\fr{\la _1}{2}\chi'^2+\fr{\la _3}{2}\chi ^2\right] \bar{h}^{kl}\bar{h}_{kl}. \label{0.31}
	\ee
For this Hamiltonian to be bounded below, the coefficients $\la _0$, $\la _1$, and $\la _3$ must be positive. 
Let us check the stability in the vicinity of the horizon where $h\simeq 0$. 
Near the horizon, one can approximate $\la _0$ and $\la _1$ 
as\footnote{Note that $X$ is finite at the horizon for the regular solution.}
	\begin{align}
	\la _0&\approx -\fr{q^2r^2}{h^2}\left( \fr{b_1}{2}-b_2\fr{fh'}{rh}\right) ,\label{0.22a} \\
	\la _1&\approx \fr{q^2r^2f}{h}\left( \fr{b_1}{2}-b_2\fr{fh'}{rh}\right) .\label{0.22b}
	\end{align}
Therefore, when the scalar velocity charge $q$ is nonzero, we get
	\be
	\la _0\la _1\approx -\fr{q^4r^4f}{h^3}\left( \fr{b_1}{2}-b_2\fr{fh'}{rh}\right) ^2<0. \label{0.23}
	\ee
This means that either $\la _0$ or $\la _1$ is negative. 
Thus, the solution given in \cite{Charmousis:2015txa} is always plagued either by the 
ghost or gradient instability.
This instability is akin to that found in \cite{Ogawa:2015pea}
for four-dimensional BHs with time-dependent scalar hair
in shift-symmetric scalar-tensor theories.

%%%%%%%%%%%%%%%%%%%%%%%%%%%%%%%%%%%%%%%%%%%%%%%%%%%%%%%%%%%%%%%%%%%%%%%%%%%%%%%%%%%%
%%%%%%%%%%%%%%%%%%%%%%%%%%%%%%%%%%%%%%%%%%%%%%%%%%%%%%%%%%%%%%%%%%%%%%%%%%%%%%%%%%%%
%	Lovelock-Galileon BH
%%%%%%%%%%%%%%%%%%%%%%%%%%%%%%%%%%%%%%%%%%%%%%%%%%%%%%%%%%%%%%%%%%%%%%%%%%%%%%%%%%%%
%%%%%%%%%%%%%%%%%%%%%%%%%%%%%%%%%%%%%%%%%%%%%%%%%%%%%%%%%%%%%%%%%%%%%%%%%%%%%%%%%%%%
\section{Lovelock-Galileon black holes in higher dimensions}\label{LGBH}
%%%%%%%%%%%%%%%%%%%%%%%%%%%%%%%%%%%%%%%%%%
%%%%%%%%%%%%%%%%%%%%%%%%%%%%%%%%%%%%%%%%%%
\subsection{Black hole solutions}

In the following, we discuss BH solutions in Lovelock-Galileon theory in arbitrary dimensions and generalize the solutions given in the previous works.\footnote{We assume that not all $b_n$'s are zero; i.e., there exists at least one nonvanishing Lovelock-derivative coupling.}
Let us consider the metric of the form
	\be
	ds^2=-h(r)dt^2+\fr{dr^2}{f(r)}+r^2\bg{ij}dx^idx^j, \label{2.1}
	\ee
where $\bg{ij}$ represents the metric of a $(D-2)$-dimensional maximally symmetric space with spatial curvature $\ka$. We also assume $\phi$ depends linearly on time as in Eq.~\eqref{0.2}:
	\be
	\phi (t,r)=qt+\psi (r). \label{2.2}
	\ee
Now we have three unknown functions of $r$:
$h(r)$, $f(r)$, and $\psi (r)$. One can show that
if we find a
solution which satisfies the $tt$-, $rr$-, and $tr$-components of Eq.~\eqref{1.7}, then it solves all the other components of Eq.~\eqref{1.7} and the scalar equation of motion \eqref{1.8}.
Thus, we have a necessary and sufficient number of equations to fully solve the system.

Let us write down these equations in terms of the unknown functions. The $tr$-component of Eq.~\eqref{1.7} is written as
	\be
	0=\sum _{n=0}^Mb_nE^{(n)}_{tr}=q\psi '\sum _{n=0}^Mb_nH^{(n)r}{}_r
	=-q\sum_{n=0}^M b_n J_r^{(n)}. \label{2.3}
	\ee
If we assume $q\psi '\ne 0$, this simplifies to
	\be
	\sum _{n=0}^M\fr{b_n}{(D-2n-1)!}F^{n-1}\left[ n\fr{fh'}{rh}-(D-2n-1)F\right] =0, \label{2.4}
	\ee
where a prime denotes a derivative with respect to $r$, and $F$ is defined by Eq.~\eqref{2.5}. Also, the $rr$-component is equivalent to
	\begin{align}
	\sum _{n=0}^M\fr{F^{n-2}}{(D-2n-1)!}\biggl\{ &n\left( a_n-\fr{b_nq^2}{2h}\right) \fr{fh'}{rh}F-(D-2n-1)\left( a_n+\fr{b_nq^2}{2h}\right) F^2 \nonumber \\
	&~~~~~+nb_nf^2\psi '^2\left[ \fr{h'}{rh}F+(D-2n-1)\fr{F}{r^2}-(n-1)\fr{fh'}{r^3h}\right] \biggr\} =0, \label{2.6}
	\end{align}
and the $tt$-component is given by
	\begin{align}
	\sum _{n=0}^M&\fr{F^{n-2}}{(D-2n-1)!}\biggl\{ \left( a_n-\fr{b_nq^2}{2h}\right) \left[ n\fr{f'}{r}F-(D-2n-1)F^2\right] +nb_n\fr{f^2}{r}F\left( \psi '^2\right) ' \nonumber \\
	&~~~~~~~~~~~~~~~+nb_nf^2\psi '^2\left[ \fr{3}{2}\fr{f'}{rf}F+\fr{1}{2}\fr{h'}{rh}F+(D-2n-1)\fr{F}{r^2}-(n-1)\fr{f'}{r^3}\right] \biggr\} =0. \label{2.7}
	\end{align}
Note that, in obtaining Eqs.~\eqref{2.6} and \eqref{2.7}, we used
Eq.~\eqref{2.3} to eliminate terms which are proportional to $\sum _{n=0}^Mb_nH^{(n)r}{}_r$. Solving Eq.~\eqref{2.4}, which can be regarded as an algebraic equation with respect to $f$ through Eq.~\eqref{2.5}, one can express $f$ in terms of $h$ and $h'$. Substituting that relation into Eq.~\eqref{2.6}, $\psi '$ can also be related to $h$ and $h'$. Then,
Eq.~\eqref{2.7} reduces to a (nonlinear) second-order differential equation
for $h(r)$. Once $h(r)$ is obtained, $f(r)$ and $\psi '(r)$ can be calculated successively. It must be noted that the roots of the algebraic equation \eqref{2.4} should be chosen so that $f$, $h$, and $\psi$ are real.

Since the original action is quite complicated, one cannot solve the field equations explicitly for the unknown functions in general.
For this reason, we shall consider some simple cases in the following.

%%%%%%%%%%%%%%%%%%%%%%%%%%%%%%%%%%%%%%%%%%
%%%%%%%%%%%%%%%%%%%%%%%%%%%%%%%%%%%%%%%%%%
\begin{enumerate}
\item {\bf Lovelock-Galileon theory of $\ell$th order}\\
As an exactly solvable example, let us focus on the
case where only the $n=\ell$ term of the Lovelock-Galileon action \eqref{1.6b} is nonzero. Namely, we consider the following action:
	\be
	S=\int d^Dx\sqrt{-g}\left[ a_\ell \mR ^{(\ell )}+b_\ell H^{(\ell )\al \beta}\phi _{;\al}\phi _{;\beta}\right] .\label{2.99}
	\ee
Since there are only trivial solutions for $\ell =0$ or $(D-1)/2$, we assume $1\le \ell <(D-1)/2$. In this case, the following nontrivial solution is obtained:
	\begin{align}
	h&=C_0-\fr{C_1}{r^{(D-2\ell -1)/\ell}}, \label{2.100a} \\
	f&=\fr{\ka}{C_0}h, \label{2.100b} \\
	\psi '^2&=\fr{q^2}{\ka h^2}\fr{C_1}{r^{(D-2\ell -1)/\ell}}, \label{2.100c}
	\end{align}
where $C_0$ and $C_1$ are constants of integration.
For this solution we find that the kinetic term of the scalar field is constant: $X=q^2/2C_0$.
Note that the above solution is independent of $a_\ell$ and $b_\ell$.
Since $\ka=0$ yields $f=0$ which is irrelevant, let us consider the $\kappa = 1$ case.
In this case one can rescale the time coordinate to have $C_0=1$ and $f=h$, leading to
	\begin{align}
	h=f&=1-\fr{C_1}{r^{(D-2\ell -1)/\ell}}, \label{2.101a} \\
	\psi '&=\pm\fr{q}{h}\fr{\sqrt{C_1}}{r^{ (D-2\ell -1)/2\ell}}. \label{2.101c}
	\end{align}
(For $\psi '$ to be real, $C_1$ must be positive.) Equation \eqref{2.101a} can be thought of as a generalization of the Schwarzschild solution in general relativity.

~~~The solution (\ref{2.101a}) has a horizon at $r=C_1^{\ell/(D-2\ell -1)}\equiv r_h$.
Clearly, $\phi=qt+\psi$ exhibits logarithmic divergence $\sim \ln |r-r_h|$ for fixed $t$.
However, following \cite{Babichev:2013cya}, by replacing $t$ in $\phi$ with the ingoing 
Eddington-Finkelstein coordinate $u$ defined by
\be
u=t+\int^r \frac{dr'}{f(r')},
\ee
we have $\phi=qu+\Psi(r)$.
It can be confirmed that $\Psi (r)$ remains finite for $r \to r_h$ for the plus branch of
Eq.~(\ref{2.101c}).
Thus, any infalling observer records a finite value of $\phi$ at the horizon $r_h$.

~~~In the case of $\kappa = -1$, one may take $C_0=-1$ and $C_1<0$, so that $h,\, f,\, \psi'^2>0$
for $0<r<r_h$.

\item {\bf Schwarzschild-like metric}\\
Even when there are a number of
nonvanishing terms in the action, one can proceed further assuming that $h(r)=f(r)$.
In this case, we can immediately integrate Eq.~\eqref{2.4} and
get the following algebraic equation for $F$:
	\be
	W[F;b_n]\equiv
	\sum _{n=0}^M\fr{b_n}{(D-2n-1)!}F^n=\fr{\mu}{r^{D-1}}, \label{2.9}
	\ee
where $\mu$ is an integration constant.
The roots should be chosen so that $F$ is real for any $r>0$.
Interestingly, with the replacement $b_n\to a_n$
the structure of Eq.~(\ref{2.9}) is identical to that of
Eq.~(\ref{defW}) obtained in the course of deriving spherically symmetric solutions
in Lovelock theory (see the Appendix). This means that
the metric functions are of the same form as those of the BH solutions in Lovelock theory.

~~~Next, we analyze Eqs.~\eqref{2.6} and \eqref{2.7} and
discuss when this type of solution is consistent and possible. Subtracting Eq.~\eqref{2.7} from Eq.~\eqref{2.6} yields
	\be
	\fr{\mu f}{r^{D+1}F'}\left( f\psi '^2-\fr{q^2}{f}\right) '=-\fr{2\mu f}{r^{D+1}F'}X'=0.\label{2.10}
	\ee
Assuming $\mu \ne 0$, we obtain $X\equiv X_0={\rm const.}$ and hence have the following relation:
	\be
	f^2\psi '^2=q^2-2X_0f. \label{2.11}
	\ee
Substituting this into Eq.~\eqref{2.6}, we obtain
	\be
	-\sum _{n=0}^M\fr{a_n-2X_0nb_n}{(D-2n-1)!}
	\left( r^{D-1}F^n\right) '+(q^2-2X_0\ka )\sum _{n=1}^M\fr{nb_n}{(D-2n-1)!}\left( r^{D-3}F^{n-1}\right) '=0. \label{2.12}
	\ee
This expression is easily integrated to give
	\be
	-\sum _{n=0}^M\fr{a_n-2X_0nb_n}{(D-2n-1)!}F^n
	+\fr{q^2-2X_0\ka}{r^2}\sum _{n=1}^M\fr{nb_n}{(D-2n-1)!}F^{n-1}
	=\fr{\nu}{r^{D-1}}, \label{2.13}
	\ee
where $\nu$ is an integration constant. Rewriting the right-hand side using Eq.~\eqref{2.9} yields
	\be
	\sum _{n=0}^M\fr{a_n-2X_0nb_n+(\nu/\mu) b_n}{(D-2n-1)!}F^n=
	\fr{q^2-2X_0\ka}{r^2}\sum _{n=1}^M\fr{nb_n}{(D-2n-1)!}F^{n-1}. \label{2.14}
	\ee
Note that $r$ in the right-hand side is related to $F$ by Eq.~\eqref{2.9}.
If $r$ is not written as a rational function of $F$, Eq.~\eqref{2.14} is satisfied if and only
if\footnote{If $r^2$ is written as a rational function of $F$, there are other possibilities.}
	\begin{align}
	q^2-2X_0\ka &=0, \label{2.15a} \\
	a_n-2X_0nb_n+(\nu/\mu) b_n&=0,~~~(0\le n\le M). \label{2.15b}
	\end{align}
%When $\ka \ne 0$ and $b_\ell \ne 0$ for some $\ell$, we can eliminate the constants $d_1$ and $\al$ to get the following $N$ conditions for $a_n$'s and $b_n$'s:
%	\be
%	\ka (a_\ell b_n-a_nb_\ell )+(n-\ell )q^2b_\ell b_n=0~~~(0\le n\le N,\; n\ne \ell ). \label{2.16}
%	\ee
From Eq.~(\ref{2.15a}) we see that the integration constant $X_0$
must be fixed as $X_0=q^2/2\kappa$ for $\kappa\neq 0$.
If $a_i=0$ (respectively $b_i=0$) then $b_i=0$ (respectively $a_i=0$)
is required from Eq.~(\ref{2.15b}). The same equation also implies that
for all nonvanishing pairs of $(a_j,\,b_j)$
\begin{eqnarray}
\frac{a_j}{b_j}-2X_0j=-\frac{\nu}{\mu}
\end{eqnarray}
must be satisfied. Thus, the solution is consistent provided that
the parameters of the theory fulfill the above requirements.
When $\ka \ne 0$, we obtain the following solution for $\psi (r)$:
	\be
	\psi '=\pm \fr{q}{f}\sqrt{1-\fr{f}{\ka}}. \label{2.17}
	\ee
Again, for the plus branch solution, it can be verified that $\phi$ remains finite 
when it is written in terms of the ingoing Eddington-Finkelstein and radial coordinates.
These results generalize those in \cite{Charmousis:2015txa}, which dealt with the $D=5$ case.
\end{enumerate}

%%%%%%%%%%%%%%%%%%%%%%%%%%%%%%%%%%%%%%%%%%
%%%%%%%%%%%%%%%%%%%%%%%%%%%%%%%%%%%%%%%%%%
\subsection{Stability analysis}

In the following we discuss the stability of the solutions above. 
We consider the tensor perturbation of the form
	\be
	\delta g_{ab}=\delta g_{ai}=0,~~~\delta g_{ij}=r^2\chi (t,r)\bh{ij}(x^k),~~~\delta \phi=0 \label{2.20}
	\ee
as we did in Sec.~\ref{CT}. 
For the nature of the tensor perturbation, we only need to consider the $ij$-components
of the field equations for the metric.
Then, the perturbed equation of motion is given by
	\be
	\sum _{n=0}^M\left( a_n\delta H^{(n)i}{}_{j}+b_n\delta E^{(n)i}{}_{j}\right) =0, \label{2.20a}
	\ee
where
	\begin{align}
	\delta H^{(n)i}{}_{j}&=\fr{1}{2}\fr{(D-4)!nF^{n-2}}{(D-2n-1)!}\bar{h}^i_j \nonumber \\
	&~~~\times \left\{ -\fr{\ddot{\chi}}{h}\left[ (n-1)\fr{f'}{r}-(D-2n-1)F\right] +f\chi''\left[ (n-1)\fr{fh'}{rh}-(D-2n-1)F\right] \right\} ,\label{dH} \\
	\delta E^{(n)i}{}_{j}&=\fr{r^2}{2}\fr{(D-4)!nF^{n-2}}{(D-2n-1)!}\bar{h}^i_j\boldsymbol{\Biggl(}\fr{\ddot{\chi}}{h}\Biggl\{ 2(n-1)\fr{f}{r}X' \nonumber \\
	&~~~~~~~~~~~~~~+X\left[ (D-2n-1)\left( 2(n-1)\fr{f}{r^2}+F\right) +(n-1)\fr{f'}{rF}\left( F-2(n-2)\fr{f}{r^2}\right) \right] \nonumber \\
	&~~~~~~~~~~~~~~-\fr{q^2}{h}\left[ (D-2n-1)\left( (n-1)\fr{f}{r^2}+F\right) -(n-1)\fr{f}{rF}\left( \fr{h'}{h}F+(n-2)\fr{f'}{r^2}\right) \right] \Biggr\} \nonumber \\
	&~~~+f\chi''\Biggl\{ X\left[ (D-2n-1)\left( 2(n-1)\fr{f}{r^2}-F\right) +(n-1)\fr{fh'}{rhF}\left( 3F-2(n-2)\fr{f}{r^2}\right) \right] \nonumber \\
	&~~~~~~~~~~~~~~-\fr{q^2}{h}\left[ (D-2n-1)\left( (n-1)\fr{f}{r^2}-F\right) +(n-1)\fr{fh'}{rhF}\left( F-(n-2)\fr{f}{r^2}\right) \right] \Biggr\} \nonumber \\
	&~~~+(\dot{\chi}'~{\rm and}~\chi~{\rm terms})\boldsymbol{\Biggr)}, \label{dE}
	\end{align}
with $X$ being the canonical kinetic term of the scalar field given by Eq.~\eqref{kin}. In Eq.~\eqref{dE}, we omitted terms containing $\dot{\chi}'$ and $\chi$ because we are only interested in the sign in front of $\ddot{\chi}$ and $\chi''$. 
From the linear differential equation for $\chi$,
we can construct the corresponding second-order action for $\chi$ as
	\be
	S^{(2)}=\int d^Dx\sqrt{-\bar{g}}\left( \fr{\la _0}{2}\dot{\chi}^2-\fr{\la _1}{2}\chi'^2+\fr{\la _2}{2}\dot{\chi}\chi'-\fr{\la _3}{2}\chi ^2\right) \bar{h}^{kl}\bar{h}_{kl}.\label{2.21}
	\ee
The coefficients of the action can be read off from the perturbed equation of motion except for an overall constant. To fix this constant, we calculate the second-order Lovelock action, which corresponds to setting $b_n=0$ in the full action. Integrating by parts, we obtain
	\begin{align}
	S_L^{(2)}&=\int d^Dx\sqrt{-\bar{g}}\sum _{n=0}^M\fr{a_n}{4}\fr{(D-4)!nF^{n-2}}{(D-2n-1)!}\bar{h}^{kl}\bar{h}_{kl} \nonumber \\
	&~~~\times \left\{ -\fr{\dot{\chi}^2}{h}\left[ (n-1)\fr{f'}{r}-(D-2n-1)F\right] +f\chi'^2\left[ (n-1)\fr{fh'}{rh}-(D-2n-1)F\right] \right\} .
	\end{align}
As a result, the full expressions for $\la _0$ and $\la _1$ are given by
	\begin{align}
	\la _0&=\sum _{n=1}^M\fr{(D-4)!nF^{n-2}}{(D-2n-1)!}\fr{r^2}{4h}\Biggl\{ -\fr{2a_n}{r^2}\left[ (n-1)\fr{f'}{r}-(D-2n-1)F\right] +4(n-1)b_n\fr{f}{r}X' \nonumber \\
	&~~~~~~~~~~+2b_nX\left[ (D-2n-1)\left( 2(n-1)\fr{f}{r^2}+F\right) +(n-1)\fr{f'}{rF}\left( F-2(n-2)\fr{f}{r^2}\right) \right] \nonumber \\
	&~~~~~~~~~~-\fr{2b_nq^2}{h}\left[ (D-2n-1)\left( (n-1)\fr{f}{r^2}+F\right) -(n-1)\fr{f}{rF}\left( \fr{h'}{h}F+(n-2)\fr{f'}{r^2}\right) \right] \Biggr\} ,\label{2.22a} \\
	\la _1&=\sum _{n=1}^M\fr{(D-4)!nF^{n-2}}{(D-2n-1)!}\fr{r^2f}{4}\Biggl\{ -\fr{2a_n}{r^2}\left[ (n-1)\fr{fh'}{rh}-(D-2n-1)F\right] \nonumber \\
	&~~~~~~~~~~+2b_nX\left[ (D-2n-1)\left( 2(n-1)\fr{f}{r^2}-F\right) +(n-1)\fr{fh'}{rhF}\left( 3F-2(n-2)\fr{f}{r^2}\right) \right] \nonumber \\
	&~~~~~~~~~~-\fr{2b_nq^2}{h}\left[ (D-2n-1)\left( (n-1)\fr{f}{r^2}-F\right) +(n-1)\fr{fh'}{rhF}\left( F-(n-2)\fr{f}{r^2}\right) \right] \Biggr\} .\label{2.22b}
	\end{align}
Now, as in Sec.~\ref{CT}, both $\la _0$ and $\la _1$ must be positive 
in order that neither ghost nor gradient instability appears. However,
near the horizon we find
	\be
	\la _0\la _1\approx -\fr{q^4r^4f}{4h^3}\left\{ \sum _{n=1}^M\fr{(D-4)!nb_nF^{n-2}}{(D-2n-1)!}\left[ (n-1)\fr{fh'}{rh}-(D-2n-1)F\right] \right\} ^2<0, \label{2.23}
	\ee
for $q\ne 0$.
Thus, we conclude that
the instability found in Sec.~\ref{CT} occurs quite generically
for hairy BHs in Lovelock-Galileon theories in $D$ dimensions.

%%%%%%%%%%%%%%%%%%%%%%%%%%%%%%%%%%%%%%%%%%%%%%%%%%%%%%%%%%%%%%%%%%%%%%%%%%%%%%%%%%%%
%%%%%%%%%%%%%%%%%%%%%%%%%%%%%%%%%%%%%%%%%%%%%%%%%%%%%%%%%%%%%%%%%%%%%%%%%%%%%%%%%%%%
%	conclusion
%%%%%%%%%%%%%%%%%%%%%%%%%%%%%%%%%%%%%%%%%%%%%%%%%%%%%%%%%%%%%%%%%%%%%%%%%%%%%%%%%%%%
%%%%%%%%%%%%%%%%%%%%%%%%%%%%%%%%%%%%%%%%%%%%%%%%%%%%%%%%%%%%%%%%%%%%%%%%%%%%%%%%%%%%
\section{Conclusions}\label{conclusion}

In the first part of this paper, we analyzed the stability of the BH solutions
in five-dimensional Lovelock-Galileon theory given in \cite{Charmousis:2015txa}.
In the analysis we calculated the second-order action and showed that the Hamiltonian cannot be bounded below
when the scalar velocity charge is nonzero.
Then we have considered the Lovelock-Galileon theory in arbitrary dimensions, which generalizes the study of \cite{Charmousis:2015txa}. 
Our ansatz is that the metric is static and spherically symmetric, and the scalar field
is allowed to contain a term linear in time in addition to the nontrivial radial profile.
We have reformulated the problem of finding the solutions into the one of solving the algebraic
equation and the second-order differential equation.
Quite remarkably, tensor perturbation analysis for the generalized solutions shows that the
Hamiltonian for the perturbation is always unbounded in the vicinity
of the horizon, exactly in the same way as in the case of five dimensions.
We thus conclude that the instability of BHs with time-dependent scalar hair in the Lovelock-Galileon theory 
is inherent in arbitrary higher dimensions. 

%%%%%%%%%%%%%%%%%%%%%%%%%%%%%%%%%%%%%%%%%%%%%%%%%%%%%%%%%%%%%%%%%%%%%%%%%%%%%%%%%%%%
%%%%%%%%%%%%%%%%%%%%%%%%%%%%%%%%%%%%%%%%%%%%%%%%%%%%%%%%%%%%%%%%%%%%%%%%%%%%%%%%%%%%

\vskip 1cm
\noindent
{\large\bf Acknowledgments}~~~~
This work was supported in part by the JSPS Grant-in-Aid for Young 
Scientists (B) Nos.~15K17632 (T.\,S.) and 24740161 (T.\,K.), MEXT Grant-in-Aid for Scientific Research on
Innovative Areas New Developments in Astrophysics
Through Multi-Messenger Observations of Gravitational
Wave Sources No.~15H00777 (T.\,S.), and MEXT KAKENHI No.~15H05888 (T.\,S. and T.\,K.).

%%%%%%%%%%%%%%%%%%%%%%%%%%%%%%%%%%%%%%%%%%%%%%%%%%%%%%%%%%%%%%%%%%%%%%%%%%%%%%%%%%%%
%%%%%%%%%%%%%%%%%%%%%%%%%%%%%%%%%%%%%%%%%%%%%%%%%%%%%%%%%%%%%%%%%%%%%%%%%%%%%%%%%%%%

\appendix
\def\thesection{Appendix:}
\section{Spherically symmetric solutions in Lovelock theory}\label{appA}

In this appendix, we summarize the basic equations
for a spherically symmetric metric in Lovelock theory \cite{Wheeler:1985qd}.
They can be reproduced simply by setting $b_n=0$ in Eqs.~(\ref{2.6}) and~(\ref{2.7}).
Obviously, the $tr$-equation is trivial in this case. We thus have
\begin{eqnarray}
\sum_{n=0}^M\frac{F^{n-1}}{(D-2n-1)!}
\left[
na_n\frac{fh'}{rh}-(D-2n-1)a_nF
\right]&=&0,
\\
\sum_{n=0}^M\frac{F^{n-1}}{(D-2n-1)!}
\left[
na_n\frac{f'}{r}-(D-2n-1)a_nF
\right]&=&0.\label{second}
\end{eqnarray}
These two equations imply that
\begin{eqnarray}
\sum_{n=0}^M\frac{na_nF^{n-1}}{(D-2n-1)!r}\left(f'-\frac{fh'}{h}\right)=0
\quad\Rightarrow\quad h\propto f.
\end{eqnarray}
One can rescale the time coordinate so that $h=f$.
Equation (\ref{second}) can be recast in
\begin{eqnarray}
\frac{d}{dr} \sum_{n=0}^M\frac{a_nr^{D-1}F^n}{(D-2n-1)!} =0,
\end{eqnarray}
and this can be integrated to give
\begin{eqnarray}
W[F;a_n]\equiv \sum_{n=0}^M\frac{a_n}{(D-2n-1)!}F^n=\frac{\mu}{r^{D-1}},\label{defW}
\end{eqnarray}
where $\mu$ is an integration constant. From
Eq.~(\ref{defW}), $F$ is determined algebraically.
In the case where $n\ge 3$ terms vanish, the well-known Boulware-Deser solution \cite{Boulware:1985wk} is recovered.

%%%%%%%%%%%%%%%%%%%%%%%%%%%%%%%%%%%%%%%%%%%%%%%%%%%%%%%%%%%%%%%%%%%%%%%%%%%%%%%%%%%%
%%%%%%%%%%%%%%%%%%%%%%%%%%%%%%%%%%%%%%%%%%%%%%%%%%%%%%%%%%%%%%%%%%%%%%%%%%%%%%%%%%%%


\begin{thebibliography}{10}
\bibitem{Lovelock-Rund}
David Lovelock and Hanno Rund,
\newblock {\em Tensors, Differential Forms, and Variational Principles}
\newblock (Dover Publications, New York, 1975).

\bibitem{Horndeski:1974wa}
Gregory~Walter Horndeski,
\newblock {Second-order scalar-tensor field equations in a four-dimensional
  space},
\newblock Int. J. Theor. Phys. {\bf 10}, 363 (1974).

\bibitem{Deffayet:2009mn}
C.~Deffayet, S.~Deser, and G.~Esposito-Farese,
\newblock {Generalized Galileons: All scalar models whose curved background
  extensions maintain second-order field equations and stress-tensors},
\newblock Phys. Rev. D {\bf 80}, 064015 (2009).

\bibitem{Deffayet:2011gz}
C.~Deffayet, Xian Gao, D.A. Steer, and G.~Zahariade,
\newblock {From $k$-essence to generalised Galileons},
\newblock Phys. Rev. D {\bf 84}, 064039 (2011).

\bibitem{Kobayashi:2011nu}
Tsutomu Kobayashi, Masahide Yamaguchi, and Jun'ichi Yokoyama,
\newblock {Generalized G-inflation: Inflation with the most general
  second-order field equations},
\newblock Prog. Theor. Phys. {\bf 126}, 511 (2011).

\bibitem{Hui:2012qt}
Lam Hui and Alberto Nicolis,
\newblock {No-Hair Theorem for the Galileon},
\newblock Phys. Rev. Lett. {\bf 110}, 241104 (2013).

\bibitem{Sotiriou:2013qea}
Thomas~P. Sotiriou and Shuang-Yong Zhou,
\newblock {Black hole hair in generalized scalar-tensor gravity},
\newblock Phys. Rev. Lett. {\bf 112}, 251102 (2014).

\bibitem{Rinaldi:2012vy}
Massimiliano Rinaldi,
\newblock {Black holes with non-minimal derivative coupling},
\newblock Phys. Rev. D {\bf 86}, 084048 (2012).

\bibitem{Babichev:2013cya}
Eugeny Babichev and Christos Charmousis,
\newblock {Dressing a black hole with a time-dependent Galileon},
\newblock J. High Energy Phys. 08 (2014) 106.

\bibitem{Anabalon:2013oea}
Andres Anabalon, Adolfo Cisterna, and Julio Oliva,
\newblock {Asymptotically locally AdS and flat black holes in Horndeski
  theory},
\newblock Phys. Rev. D {\bf 89}, 084050 (2014).

\bibitem{Minamitsuji:2013ura}
Masato Minamitsuji,
\newblock {Solutions in the scalar-tensor theory with nonminimal derivative
  coupling},
\newblock Phys. Rev. D {\bf 89}, 064017 (2014).

\bibitem{Bravo-Gaete:2013dca}
Moises Bravo-Gaete and Mokhtar Hassaine,
\newblock {Lifshitz black holes with a time-dependent scalar field in a
  Horndeski theory},
\newblock Phys. Rev. D {\bf 89}, 104028 (2014).

\bibitem{Cisterna:2014nua}
Adolfo Cisterna and Cristi$\acute{\rm a}$n Erices,
\newblock {Asymptotically locally AdS and flat black holes in the presence of
  an electric field in the Horndeski scenario},
\newblock Phys. Rev. D {\bf 89}, 084038 (2014).

\bibitem{Kobayashi:2014eva}
Tsutomu Kobayashi and Norihiro Tanahashi,
\newblock {Exact black hole solutions in shift symmetric scalar-tensor
  theories},
\newblock Prog. Theor. Exp. Phys. {\bf 2014}, 073E02 (2014).

\bibitem{Lovelock:1971yv}
D.~Lovelock,
\newblock {The Einstein tensor and its generalizations},
\newblock J. Math. Phys. {\bf 12}, 498 (1971).

\bibitem{Charmousis:2015txa}
Christos Charmousis and Minas Tsoukalas,
\newblock {Lovelock Galileons and black holes},
\newblock Phys. Rev. D {\bf 92}, 104050 (2015).

\bibitem{Ogawa:2015pea}
Hiromu Ogawa, Tsutomu Kobayashi, and Teruaki Suyama,
\newblock {Instability of hairy black holes in shift-symmetric Horndeski
  theories},
\newblock arXiv:1510.07400.

\bibitem{Higuchi:1986wu}
Atsushi Higuchi,
\newblock {Symmetric tensor spherical harmonics on the $N$ sphere and their
  application to the de Sitter group $SO$($N$,1)},
\newblock J. Math. Phys. {\bf 28}, 1553 (1987);
\newblock {\bf 43}, 6385(E) (2002).

\bibitem{Wheeler:1985qd}
James~T. Wheeler,
\newblock {Symmetric Solutions to the Maximally {Gauss-Bonnet} Extended
  Einstein Equations},
\newblock Nucl. Phys. {\bf B273}, 732 (1986).

\bibitem{Boulware:1985wk}
David~G. Boulware and Stanley Deser,
\newblock {String Generated Gravity Models},
\newblock Phys. Rev. Lett. {\bf 55}, 2656 (1985).
\end{thebibliography}
\end{document}